\documentclass[11pt,a4paper]{article}
\pdfoutput=1 
\usepackage[utf8]{inputenc}
\usepackage{jheppub}
\usepackage{graphicx}
\usepackage{caption}
\usepackage{subcaption}
\usepackage{hyperref}
\usepackage[T1]{fontenc}
\usepackage{makecell}

\newcommand{\be}{\begin{equation}}
\newcommand{\ee}{\end{equation}}
\newcommand{\bea}{\begin{eqnarray}}
\newcommand{\eea}{\end{eqnarray}}

\def\CC{\mathcal{C}}

\def\CE{\mathcal{E}}

\def\CT{\mathcal{T}}

\title{A family of three maximally symmetric boost-invariant flows in relativistic hydrodynamics}

\author[a,b]{Sa\v{s}o Grozdanov}
\affiliation[a]{Higgs Centre for Theoretical Physics,  School of Physics and Astronomy, University of Edinburgh, \\Edinburgh, EH8 9YL, Scotland}
\affiliation[b]{Faculty of Mathematics and Physics, University of Ljubljana, \\Jadranska ulica 19, SI-1000 Ljubljana, Slovenia}

\abstract{I discuss the constructions of boost-invariant dissipative conformal hydrodynamic flows by elaborating on the geometric procedure by Gubser and Yarom, which starts from a static, maximally symmetric flow on dS${}_3\times\mathbb{R}$. Three foliations of dS$_3$ preserve three-dimensional non-Abelian isometry groups, namely, the flat ISO(2)-invariant, the spherical (closed) SO(3)-invariant, and the hyperbolic (open) SO(2,1)-invariant slicings. I show that the fluids that preserve these symmetries, after they have been Weyl transformed to flat spacetime, give rise to three physically distinct and boost-invariant solutions of the relativistic dissipative Navier-Stokes equations: the well-known and widely studied Bjorken and Gubser flows, and a seemingly thus far unexplored solution that arises from the hyperbolic slicing of dS${}_3$. The new solution combines the radial expansion characteristic of the Gubser flow with the late-proper-time applicability of Bjorken’s solution, and features a finite, radially bounded droplet whose expanding edge resembles a free-streaming shockwave.}

\begin{document} 
\maketitle
\flushbottom

\section{Introduction}

Exact analytic solutions of non-linear partial differential equations are rare. This is especially true if the said solutions are meant to model interesting physical processes. The Navier-Stokes equations that describe the dynamics of dissipative fluids are no exception. These equations, which express the near-equilibrium evolution of long-lived conserved quantities, such as energy and momentum, have for over two centuries been successfully applied to an incredible array of phenomena in gases and liquids (see e.g.~Refs.~\cite{landau,Kovtun:2012rj,Romatschke:2017ejr}). What may seem more surprising is that these equations in their relativistic form also describe certain phenomena in high-energy particle physics. In particular, they have been successfully applied to the description of heavy-ion collisions and the evolution of the quark-gluon plasma (QGP) (see Ref.~\cite{Busza:2018rrf}). Even though such phenomena occur at energies above a few~TeV per nucleon pair 
(e.g.,~$\sqrt{s_{NN}} \sim 2.8$--$5.0~\mathrm{TeV}$ at the LHC) and the size of a typical 
QGP droplet is only of order a few femtometers ($\sim 5$--$10~\mathrm{fm}$), 
due to the strong interactions of quantum chromodynamics (QCD), such states 
hydrodynamise in a remarkably short time (see Refs.~\cite{Chesler:2013lia,vanderSchee:2013pia,Heller:2016gbp,Grozdanov:2016zjj}). 

It was in 1982 that Bjorken found an exact 4$d$ boost-invariant solution of relativistic and conformal Navier-Stokes equations \cite{Bjorken:1982qr}. This solution has been used extensively in modeling of high-energy flows, and has been applied to the physics of heavy ions. It enjoys the following set of symmetries: SO(1,1)$\times$ISO(2)$\times\mathbb{Z}_2$. Respectively, they correspond to boost invariance along the beam, translations and rotations in the transverse plane (i.e., the isometries of the 2$d$ Euclidean plane) and reflection symmetry along the beam. 

Another boost-invariant conformal solution was later found by Gubser in 2010 \cite{Gubser:2010ze}. Its symmetry algebra has the same number of dimensions and is given by SO(1,1)$\times$SO(3)$ \times\mathbb{Z}_2$, where SO(3) is a subgroup of the conformal group SO(4,2). This group structure was elucidated from a geometric point of view by Gubser and Yarom in Ref.~\cite{Gubser:2010ui}.\footnote{This method as well as other methods have been used to find other exact solutions, including some with rotation and some that include effects from second-order hydrodynamics \cite{Hatta:2014gqa,Hatta:2014gga,Nagy:2007xn,Nagy:2009eq,Bantilan:2018vjv,Csorgo:2018pxh,Rodgers:2022fuv}.} While the Bjorken flow is translationally invariant in the transverse plane, the main advantage of the Gubser flow is its expanding radial profile. 

In this note, I reexamine the symmetry structure of the Bjorken and Gubser solutions, and show that another, third solution exists in the same class of conformal boost-invariant flows with the same number of symmetry generators. Namely, instead of ISO(2) or SO(3), a fluid can be invariant under SO(2,1). In the geometric picture, these three symmetry groups correspond to the flat, closed spherical, and open hyperbolic slicings of the auxiliary 3$d$ de Sitter space used in the construction by Ref.~\cite{Gubser:2010ui}. From this point of view, the Gubser flow is not a generalisation of the Bjorken flow. Both of those solutions, as well as the new one presented here, should be seen as three equivalent members of the same `maximally symmetric' family of viscous boost-invariant solutions of the relativistic dissipative Navier-Stokes equations.

\section{Construction of the three boost-invariant flows}\label{Sec:Construction}

I begin by constructing the boost-invariant flows with `maximal' symmetry from a single, unified procedure that follows that of Gubser and Yarom \cite{Gubser:2010ui}. The main strength of the approach was in providing a clear geometric understanding of the symmetries that underlie Gubser's solution \cite{Gubser:2010ze} known as the Gubser flow. Here, I show that there exist three such solutions, which are all symmetric under non-trivial symmetry algebras with the same dimension.

\subsection{Ideal hydrodynamic solutions in the auxiliary (dS${}_3 \times \mathbb{R}$) space}
Consider an ideal conformal relativistic fluid in the dS${}_3 \times \mathbb{R}$ geometry, where dS${}_3$ is the three-dimensional de Sitter space. This manifold can be obtained as a Weyl transformation of the flat Minkowski space.\footnote{While we could consider other spaces that are Weyl transforms of Minkowski spacetime as the starting point (see e.g.~Ref.~\cite{Hatta:2014gga}), dS${}_3 \times \mathbb{R}$ yields the three boost-invariant solutions of interest to this paper.} There are three ways in which dS${}_3$ can be foliated with 2$d$ slices that preserve the maximal amount of symmetries, namely, a three-dimensional algebra. These are the flat ($\kappa = 0$), the spherical ($\kappa = +1$) and the hyperbolic ($\kappa = -1$) slicings. In compact notation, setting the dS${}_3$ radius to one, we can write the metric of dS${}_3 \times \mathbb{R}$ as
\begin{equation}
    d\hat s^2_\kappa = - d\rho^2 + S^2_\kappa(\rho) \left( d\theta^2 + Q^2_\kappa(\theta) d\phi^2 \right) + d\eta^2  ,
\end{equation}
where $(\rho, \theta, \phi)$ parametrise dS${}_3$ and $\eta\in(-\infty,\infty)$ parametrises $\mathbb{R}$. In the three cases, 
\begin{align}
    \kappa = 0:&  &S_0(\rho) &= e^{-\rho}, &  Q_0(\theta) &= \theta,   \\
    \kappa = +1:& & S_1(\rho) &= \cosh\rho,  & Q_1(\theta) &= \sin\theta, \\
    \kappa = -1:& & S_{-1}(\rho) &= \sinh\rho,  & Q_{-1}(\theta) &= \sinh\theta,
\end{align}
with the ranges of the coordinates on dS${}_3$ given by
\begin{align}
    \kappa = 0:&  & -\infty &< \rho < \infty, && 0 \leq \theta < \infty,& &0\leq \phi < 2 \pi ,  \\
    \kappa = +1:& &  -\infty &< \rho < \infty,&& 0 \leq \theta < 2\pi, & &0\leq\phi<2\pi, \\
    \kappa = -1:& & 0 &< \rho < \infty, & &0 \leq \theta < \infty, && 0 \leq \phi < 2\pi.
\end{align}
The three-dimensional isometry groups that correspond to the $(\theta,\phi)$ slices for $\kappa = 0,\, +1,\,-1$ are ISO(2), SO(3) and SO(2,1), respectively. Note that, locally, the SO(2,1) slicing is an analytic continuation of the spherical SO(3) foliation. Arriving at the flat slicing from, e.g., the closed spherical slicing, requires a combination of operator rescalings and a limit, or, in the algebraic language, an operation called the group contraction. 

An ideal conformal fluid in its rest frame that preserves the isometries of the slicings has the energy-momentum tensor
\begin{equation}
    \hat T^{\mu\nu}_{(\kappa)} = \left(\hat\varepsilon_\kappa(\rho) + \hat p_\kappa (\rho) \right) \hat u^\mu \hat u^\nu  + \hat p_\kappa(\rho) \hat g^{\mu\nu} , \label{Ideal_Tmunu}
\end{equation}
with the velocity field $\hat u^\mu = (1,0,0,0)$ and the equation of state $\hat p_\kappa = \hat\varepsilon_\kappa/3$. Solving the equations of motion (the energy-momentum tensor conservation or the Ward identity),
\begin{equation}
    \hat\nabla_\mu \hat T^{\mu\nu}_{(\kappa)} = 0 , \label{NSEq}
\end{equation}
then gives the $\rho$-dependent energy densities that correspond to the three solutions:
\begin{align}
    \hat\varepsilon_0(\rho) = \CC_0 e^{8\rho/3}, \quad \hat\varepsilon_{+1}(\rho) = \frac{\CC_{+1}}{\cosh^{8/3}\rho}, \quad \hat\varepsilon_{-1}(\rho) = \frac{\CC_{-1}}{\sinh^{8/3}\rho}, 
\end{align}
each with an integration constant $\CC_\kappa$. All other thermodynamic scalars of this conformal (scale-free) solution can be easily worked out by using thermodynamic identities and dimensional analysis. Note that for $\kappa = \{0,+1,-1\}$, $\hat\nabla_\mu \hat u^\mu = \{-2, 2\tanh\rho, 2\coth \rho \}$. Hence, for $\kappa = +1$, $-2 < \hat\nabla_\mu \hat u^\mu < 2$, and for $\kappa = -1$, $\hat\nabla_\mu \hat u^\mu > 2$.

\subsection{Ideal hydrodynamic solutions in the physical (Minkowski) space}
To find the corresponding conformal boost-invariant flows in the physical, flat Minkowski space expressed in terms of the Milne coordinates, we perform the Weyl transformation:
\begin{equation}
    ds^2= - d\tau^2 + dr^2 + r^2 d\phi^2 + \tau^2 d\eta^2 = \tau^2 d\hat s^2_\kappa,
\end{equation}
and use the following coordinate transformations:
\begin{align}
    \kappa = 0:&  &  e^{\rho} &= \tau , &   \theta&= r ,   \\
    \kappa = +1:& &  \sinh \rho &= - \frac{1 - q^2 \left(\tau^2 - r^2\right)}{2q\tau} ,  &  \tan \theta &= \frac{2qr}{1 + q^2 \left(\tau^2 - r^2\right)} , \\
    \kappa = -1:& &  \cosh\rho &= \frac{1 + q^2 \left(\tau^2 - r^2 \right)}{2q\tau} ,  &  \tanh \theta &= -\frac{2 q r}{1 - q^2 \left(\tau^2 - r^2\right)},
\end{align}
where $q$ is a free non-negative parameter. Note that in flat space, the operators of the SO(3) (the standard Gubser flow case) and SO(2,1) algebras that fix the velocity field depend on this parameter. We could therefore also call them SO(3)${}_q$ and SO(2,1)${}_q$. In the three cases, the allowed ranges of the flat space coordinates are $0 < r < \infty$ for the coordinate perpendicular to the beam, $0 \leq \phi < 2 \pi$ for the angle in the plane transverse to the beam,
$-\infty < \eta < \infty$ along the beam, and for the proper time, 
\begin{align}
    \kappa = 0:&  \quad 0 < \tau <\infty,  \\
    \kappa = +1:& \quad 0 < \tau < \infty, \\
    \kappa = -1:& \quad \frac{1}{q} < \tau - r . \label{hyper_coordinates}
\end{align}
Importantly, the $\kappa = -1$ solution can only exist within the future lightcone originating at $r=0$, $\tau = 1/q$. The physical energy densities of the three flows are then
\begin{align}
    \varepsilon_0 (\tau) &= \frac{\CC_0}{\tau^{4/3}}, \\
    \varepsilon_{+1} (\tau, r) &= \frac{(2q)^{8/3} \CC_{+1}}{\tau^{4/3} \left[ 1 + 2q^2 (\tau^2 + r^2) + q^4 (\tau^2-r^2)^2 \right]^{4/3}}, \\
    \varepsilon_{-1}(\tau,r) &= \frac{(2q)^{8/3} \CC_{-1}}{\tau^{4/3} \left[ 1 - 2q^2 (\tau^2 + r^2) + q^4 (\tau^2-r^2)^2 \right]^{4/3}}.
\end{align}
The three flows correspond to the Bjorken flow ($\kappa = 0$), the Gubser flow ($\kappa = +1$) and a new solution with $\kappa = -1$. In compact notation, the three energy densities can be written (with the numerator constants absorbed into the constant $\CE$) as
\begin{equation}
    \varepsilon_{\kappa}(\tau,r) = \frac{\CE_{\kappa}}{\tau^{4/3} \left[ 1 + 2 \kappa q^2 (\tau^2 + r^2) + \kappa^2 q^4 (\tau^2-r^2)^2 \right]^{4/3}}.
\end{equation}
The four-velocity $u^\mu$ in Minkowski space is computed by using $u_\mu = \tau \hat u_\nu \partial \hat x^\nu/\partial x^\mu$. In a similarly unified notation, for all three $\kappa$,
\begin{align}
    u^\tau &= \frac{\kappa^2 q^2 (\tau^2 + r^2) + \kappa + (1-\kappa)(1+\kappa)}{\sqrt{1 + 2 \kappa q^2 (\tau^2 + r^2) + \kappa^2 q^4 (\tau^2-r^2)^2}}, \\
    u^r &= \frac{2 \kappa^2 q^2 \tau r}{\sqrt{1 + 2 \kappa q^2 (\tau^2 + r^2) + \kappa^2 q^4 (\tau^2-r^2)^2}}, \\
    u^\phi &= u^\eta = 0.
\end{align}
It is easy to check that $u^\mu u_\mu = -1$. Finally, we can also compute the transverse velocity:
\begin{equation}
    v_\perp = \frac{u^r}{u^\tau} = \frac{2 \kappa^2 q^2 \tau r}{\kappa^2 q^2 (\tau^2 + r^2) + \kappa + (1-\kappa)(1+\kappa)}.
\end{equation}
It is important to note that the $\kappa = -1$ solution has a singularity at the edge of its allowed coordinate range \eqref{hyper_coordinates}. It runs along the entire edge of the lightcone, $r = \tau - 1/q$, where both the ideal energy density and the velocity field diverge. 

\subsection{Viscous corrections}
Next, let us add the viscous corrections to the flows. For a conformal fluid, the only first-order hydrodynamic contribution arises due to the shear viscosity $\eta = H_0 \varepsilon^{3/4}$, where $H_0$ is a constant. In the auxiliary dS${}_3\times\mathbb{R}$, we change the energy-momentum tensor \eqref{Ideal_Tmunu} to become
\begin{equation}
    \hat T^{\mu\nu}_{(\kappa)} = \left(\hat\varepsilon_\kappa(\rho) + \hat p_\kappa (\rho) \right) \hat u^\mu \hat u^\nu  + \hat p_\kappa(\rho) \hat g^{\mu\nu} - 2 H_0 \hat\varepsilon^{3/4}_\kappa(\rho) \hat\sigma^{\mu\nu}, 
\end{equation}
where $\hat\sigma^{\mu\nu}$ is the standard symmetric, transverse and traceless tensor made of a single (covariant) derivative $\hat\nabla_\nu$ of $\hat u^\mu$ (see Refs.~\cite{Baier:2007ix,Grozdanov:2015kqa,Gubser:2010ui}). By the symmetry constraints, the velocity flow remains unchanged. The relativistic Navier-Stokes equations \eqref{NSEq} are then solved by incorporating the dissipative viscous correction into the energy density, or, more conveniently, the temperature field, where $\hat\varepsilon_\kappa = \hat T_\kappa^4$. Finally, by Weyl and coordinate transforming to flat spacetime, we obtain the solution for the Bjorken flow ($\kappa = 0$), 
\begin{equation}
    T_0(\tau) = \frac{\CT_0}{\tau^{1/3}} - \frac{H_0}{2\tau} , \label{T_B}
\end{equation}
as well as for the Gubser flow ($\kappa = +1$),
\begin{align}\label{T_G}
    T_{+1}(\tau,r) = \frac{1}{\tau \left(g^2 + 1\right)^{1/3}} \left[ \CT_{+1} - \frac{H_0 g^3}{9} {}_2 F_1\left(\frac{3}{2},\frac{7}{6};\frac{5}{2}; - g^2 \right)  \right],
\end{align}
with
\begin{equation}
    g = \frac{1 - q^2 \left(\tau^2 - r^2 \right)}{2 q \tau}.
\end{equation}
Finally, the viscous temperature field of the new solution with $\kappa = -1$ is given by\footnote{Note that among many ways of writing the viscous term, we can also use the following identity: $h^3 {}_2F_1\left(\frac{4}{3},1;\frac{5}{6}; - \left(h^2-1\right) \right) =  {}_2F_1\left(-\frac{1}{2},-\frac{1}{6};\frac{5}{6}; - \left(h^2-1\right) \right) $.}
\begin{align}\label{T_H}
    T_{-1}(\tau,r) = \frac{1}{\tau \left(h^2 - 1\right)^{1/3}}\left[\CT_{-1}  - \frac{H_0 h^3}{\left(h^2-1\right)^{1/6}} {}_2F_1\left(\frac{4}{3},1;\frac{5}{6}; - \left(h^2-1\right) \right)   \right],
\end{align}
where
\begin{equation}
    h = \frac{1+q^2\left(\tau^2 - r^2\right)}{2q\tau} .
\end{equation}
In all three cases, $\CT_\kappa$ are arbitrary constants.

\section{Discussion of the flow solutions and future directions}

I now discuss the qualitative phenomenological differences between the three boost-invariant flat-space solutions constructed in Section~\ref{Sec:Construction}. 

The Bjorken flow with $\kappa = 0$ has a constant velocity field $u^\mu = (1,0,0,0)$, which implies an absence of any transverse evolution. For small $\tau$, $T_0$ may become negative, making the solution unphysical. This occurs at $\tau = (H_0/ 2 \CT_0)^{3/2}$. We can also introduce a measure that tests the validity of the gradient expansion, i.e., a `Knudsen'-like number, $\left|\partial_\tau \ln T_0 \right|/T_0$. It varies rapidly and has cusps for small $\tau$. This is in the regime where the temperature can also be negative. Hence, the Bjorken flow is a good dissipative hydrodynamic solution in the regime of large $\tau$, consistent with typical hydrodynamic expectations (despite the fact that $\tau \to \infty$ takes $T_0 \to 0$). Since this flow has been thoroughly studied in the literature, I will now mostly focus on comparing the other two solutions.

\begin{figure}[th!]
    \centering
    \includegraphics[width=0.9\textwidth]{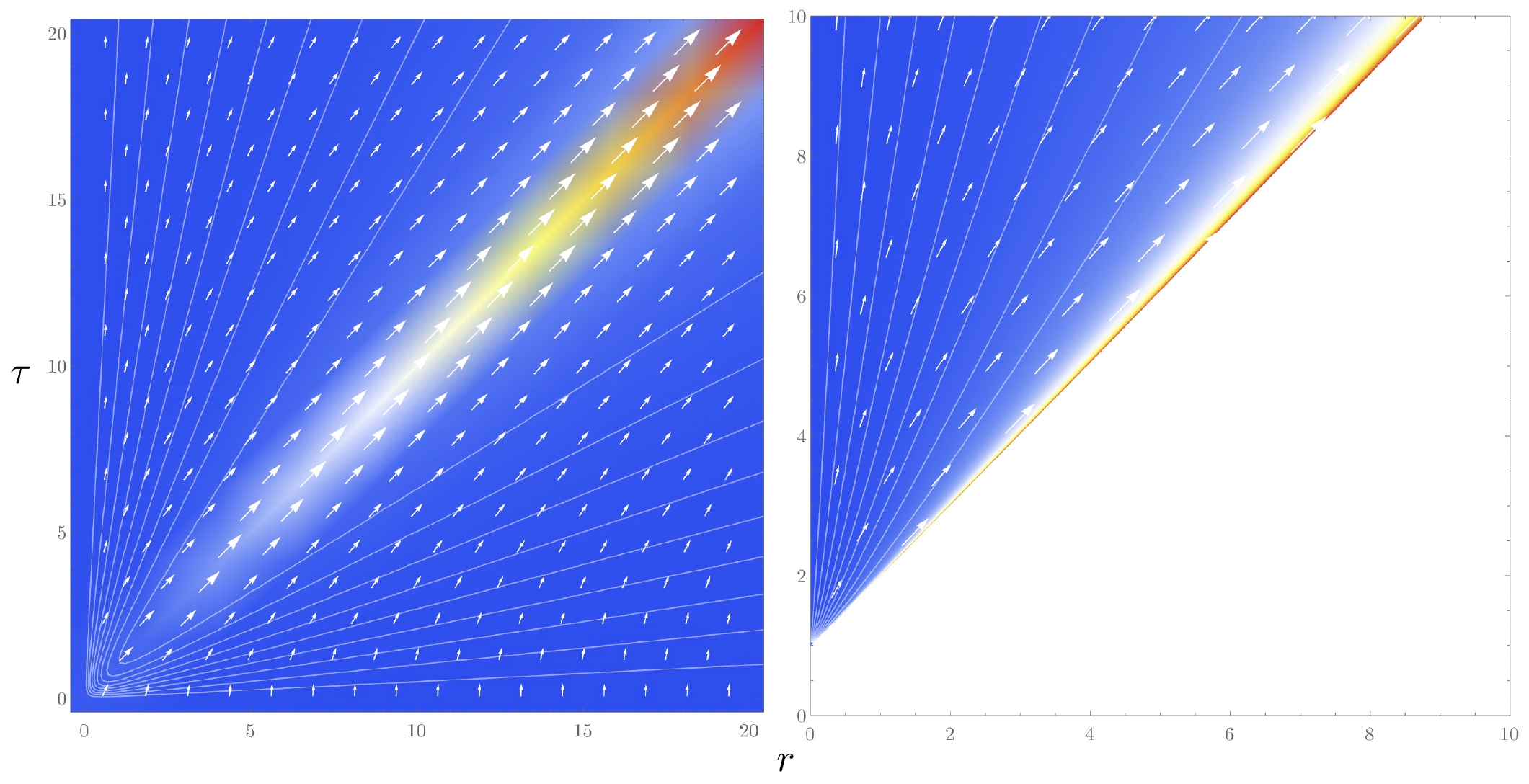}
    \caption{Velocity vector $(u^r,u^\tau)$ of the Gubser's $\kappa = +1$ ({\bf left panel}) and the $\kappa = -1$ ({\bf right panel}) solutions shown in the $(r,\tau)$ plane. Colours indicate the norm of the velocity vector, with its magnitude increasing from blue to red. White arrows show the size and the direction of $u^\mu$. White lines show contours of equal transverse $v_\perp$, running from 0 to 1 in steps of 0.1. For the Gubser flow ($\kappa = +1$), $q=2$, and for the $\kappa = -1$ flow, $q=1$. The latter solution exists inside the $r < \tau - 1/q$ future lightcone, with $u^\mu$ diverging at the edge.} 
    \label{fig:Velocity_profiles}
\end{figure}

I first compare the velocity field of the Gubser flow with the $\kappa = -1$ solution. A typical situation is depicted in Figure~\ref{fig:Velocity_profiles}. The main difference is that the Gubser flow solution is smooth and exists everywhere along the transverse plane whereas the $\kappa = -1$ solution has a sharp cutoff at the edge of the causal cone where the velocity field diverges. While the Gubser flow's magnitude of the velocity also increases near the lightcone, this rise is gradual and smooth. One implication is therefore that far from the beam, the new solution approaches free streaming of (massless, conformal) relativistic particles traveling at the speed of light, and behaves as a smooth shockwave. In both cases, the free parameter $q$ effectively sets the size of the droplet. In Gubser's everywhere-smooth solution, it sets the approximate scale of its size, whereas for $\kappa = -1$, it sets the sharp edge of the lightcone where the solution ends in a divergence. As a result, the radial expansion of the droplet that follows a `collision' of nuclei starts at the value of the proper time set by $q$, at $\tau = 1/q$.

\begin{figure}[h!]
    \centering
    \includegraphics[width=0.9\textwidth]{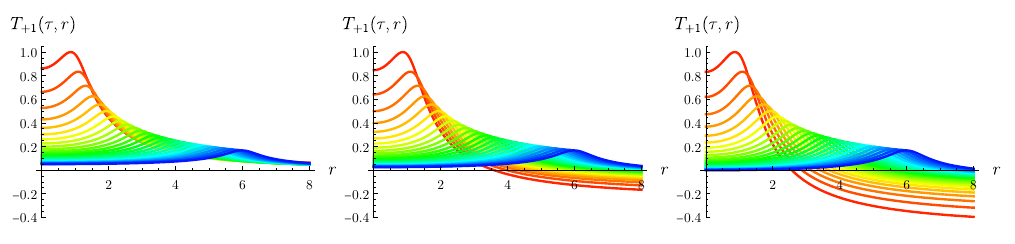}
    \includegraphics[width=0.9\textwidth]{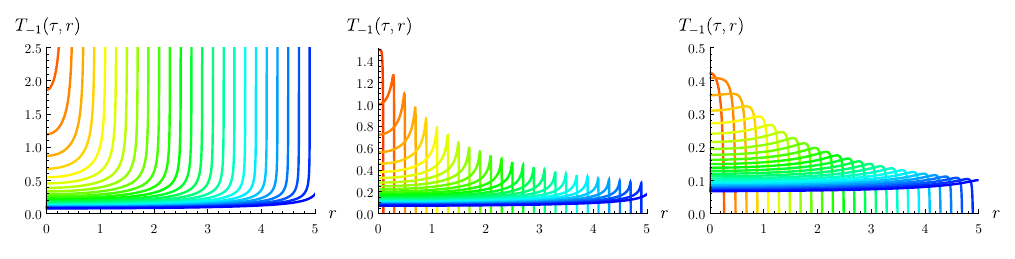}
    \caption{Evolution of the radial temperature profiles for the $\kappa = +1$ (Gubser flow) and the $\kappa = -1$ solutions. {\bf Top:} $T_{+1}$ from Eq.~\eqref{T_G} is plotted for increasing $\tau$ from $1$ (red) to $6$ (blue) in spacings of $0.2$. From left to right, the three plots show different values of the shear viscosity parameter $H_0 = \{0, 0.5, 1\}$, with the remaining parameters set to $\CT_{+1} = 1$ and $q=2$. {\bf Bottom:} $T_{-1}$ from Eq.~\eqref{T_H}, with $\tau$ running from $1.1$ (red) to $6.1$ (blue) in spacings of $0.2$. From left to right, $H_0 = \{0, 0.3, 0.5\}$, with $\CT_{-1} = 1$ and $q=1$.} 
    \label{fig:G_H_T_profiles}
\end{figure}

Differences between the two solutions can also be seen from the radial profiles of the temperature fields $T_{\pm 1}$. This is shown in Figure~\ref{fig:G_H_T_profiles}. Focusing on the ideal limit $(H_0 = 0)$, the Gubser flow is smooth for all $r$ with the magnitude of the peak decreasing with $\tau$. When viscosity is non-zero, however, the solution develops unphysical regions with $T_1 < 0$ for large $r$ at early proper time. This is unlike in the $\kappa = -1$ solution, which is confined to the lightcone. There, at $H_0 = 0$, the temperature simply diverges at the lightcone. Then, as the viscosity is turned on, this divergence is smoothened and its effect is to invalidate the solution near the lightcone by making $T_{-1} < 0$, where one could imagine cutting off the solution. In fact, the region around $T_{-1} \approx 0$, near the edge of the lightcone, is the natural place to introduce a surface on which to impose freeze-out conditions. Away from the near-lightcone region, the temperature field is positive and smooth for all $r$ and $\tau$, even at early times. This is unlike the Gubser flow which cannot be naturally cut off at a finite $r$. That solution extends to infinity in perpendicular directions to the beam, for all $\tau$.

Another important difference between the viscous flows arises in the large-$\tau$ regime. A simple way to see this and to anticipate potential pathologies is by expanding the three temperature fields. To leading order in the $\tau\to\infty$ limit, the Bjorken flow is dominated by ideal hydrodynamics, $T_0 \sim \CT_0/\tau^{1/3}$. On the other hand, the $\kappa = \pm 1$ solutions are controlled by the viscous term, $T_{\pm 1} \sim \mp H_0 / 2 \tau$. This makes it clear that the dominant term of the Gubser flow is negative, thereby implying a large-$\tau$ pathology. The $\kappa = -1$ solution, however, retains a positive decaying temperature as $\tau\to\infty$. Moreover, the Gubser flow solution only has small gradients for intermediate proper time scales, not at late $\tau$. On the other hand, the $\kappa = -1$ solution retains small temperature gradients for late $\tau$, so long as it is probed away from the null boundaries of the lightcone. This is because the energy gets accumulated near the lightcone, which allows for the relaxation of the flow everywhere inside the droplet. It is important to note, however, that despite the temperature having small and well-behaved derivatives for $\tau \gg 1$, the eventual dominance of the viscous term over the ideal term as $\tau \to \infty$ means that, strictly, the $\kappa = -1$ solution also violates the order-by-order gradient expansion. That this can happen is not particularly surprising as, indeed, all three boost-invariant flows discussed here are solutions of the fully non-linear viscous Navier-Stokes equations. I show the behaviour of the derivatives of the temperature field by plotting two representative examples of another `Knudsen'-like number defined in the units of the `curvature parameter' $q$, $\tilde K_{\pm 1} \equiv (1/q) \sqrt{ (\partial_\tau \ln T_{\pm 1})^2 + (\partial_r \ln T_{\pm 1})^2}$. The flows can be considered to be `well-behaved' for $\tilde K_\pm \lesssim 1$. 

\begin{figure}[h!]
    \centering
    \includegraphics[width=0.9\textwidth]{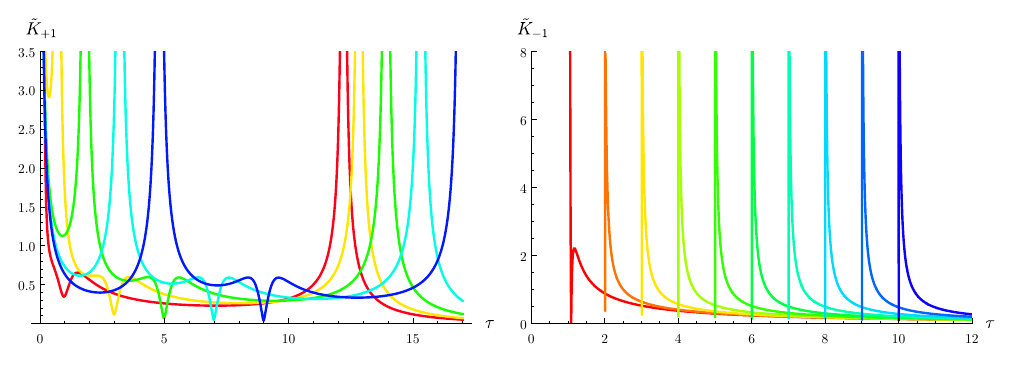}
    \caption{Plots of $\tilde K_{\pm 1} $ that quantifies the behaviour of the temperature gradients. {\bf Left:} Gubser's $\kappa = +1$ solution with $\CT_{+1} = 1$, $q=2$ and $H_0 = 0.5$, plotted as a function of $\tau$, for $r$ running from 1 (red) to 9 (blue) in steps of 2. {\bf Right:} The $\kappa = -1$  solution with $\CT_{-1} = 1$, $q=1$ and $H_0 = 0.3$, shown for $r$ between 0 (red) to 9 (blue) in steps of 1.} 
    \label{fig:G_H_Kn}
\end{figure}

In summary, the new $\kappa = -1$ solution should be phenomenologically interpreted as describing a relativistic conformal droplet confined to the future lightcone, which, as it propagates radially outwards from the beam, behaves as a free-streaming-like state at its causal edge. Similarly to the Gubser flow, the solution is boost-invariant (along the beam) and radially expanding. On the other hand, what is very different from the Gubser flow is that the solution exhibits a shockwave-like expansion at the lightcone. Rather speculatively, such behaviour suggests potential relevance of the solution for the phenomenology of the colour-glass condensate and the glasma, perhaps in connection with their nearly free-streaming early-time dynamics (see Refs.~\cite{Iancu:2003xm,Gelis:2012ri} and also Ref.~\cite{Gubser:2012gy}). Less speculatively, the $\kappa = -1$ solution may also provide a causal, boost-invariant model of a finite plasma droplet that describes the transition from hydrodynamic to ballistic expansion in heavy-ion collisions.\footnote{For an intriguing hydrodynamic--ballistic matching in unitary Fermi gases, see Refs.~\cite{Bluhm:2015bzi,Bluhm:2017rnf}.} While, as is usual with such solutions, the flow is outside of the validity of the hydrodynamic approximation at the earliest, far-from-equilibrium stage of the evolution, it should offer a realistic analytic framework for studying late-time transverse dynamics, freeze-out conditions, and for benchmarking numerical hydrodynamic simulations with finite spatial support. Therefore, this solution may naturally apply to the intermediate and late-$\tau$ regimes of a rapidly expanding quark-gluon plasma, as it retains small gradients (for any fixed $r$ away from the causal edge of the null lightcone) and thus remains at least approximately consistent with the assumptions of the hydrodynamic derivative expansion. In this sense, the new solution combines the radial expansion characteristic of the Gubser flow with the late-$\tau$ applicability of Bjorken's solution, while introducing the novel feature of a droplet with finite radial extent whose boundary expands in a manner reminiscent of a free-streaming shockwave.

Future studies should examine the solution more closely from the points of view of the M\"{u}ller-Israel-Stewart theory and the Boltzmann equation, following, e.g., the past works in Refs.~\cite{Baier:2007ix,Bemfica:2017wps,Denicol:2014tha,Nopoush:2014qba,Martinez:2017ibh}, as well as with the help of the AdS/CFT correspondence, following Refs.~\cite{Janik:2005zt,Janik:2006ft,Banerjee:2023djb,Mitra:2024zfy}. With further insights drawn from all of these perspectives, we should then be able to understand the potential for applying the $\kappa = -1$ solution to the phenomenological description of realistic high-energy processes, in particular, to the collisions of heavy ions, and the formation of quark-gluon plasma. 

\acknowledgments{I would like to thank Mauricio Martinez, Wilke van der Schee and Alexander Soloviev for helpful and stimulating discussions. This work was supported by the STFC Ernest Rutherford Fellowship ST/T00388X/1. It was also supported by the research programme P1-0402 and the project J7-60121 of Slovenian Research Agency (ARIS).}

\bibliographystyle{JHEP}
\bibliography{Genbib}{}

\end{document}